# Synergetic effect of edge states and point defects to tune ferromagnetism in CVD-grown vertical nanostructured MoS$_2$: A correlation between electronic structure and theoretical study

*Sharmistha Dey[a], Pankaj Srivastava[a], Ankita Phutela[b], Saswata Bhattacharya[b], Fouran Singh[c], and Santanu Ghosh[a*]*

[a] Nanostech Laboratory, Department of Physics, Indian Institute of Technology Delhi, New Delhi 110016, India

[b] Department of Physics, Indian Institute of Technology Delhi, New Delhi 110016, India

[c] Materials Science Group, Inter-University Accelerator Centre, New Delhi 110067, India

**Abstract:**

Room-temperature ferromagnetism (RTFM) exhibited by nanostructured two-dimensional semiconductors for spintronics applications is a fascinating area of research. The present work reports on the correlation between the electronic structure and magnetic properties of defect-engineered nano-structured MoS$_2$ thin films. Low-energy light and heavy-mass ion irradiation have been performed to create defects and tune magnetic properties. Vertical nanosheets with edge state termination in the pristine sample have been examined by field emission scanning electron microscopy (FESEM). Deterioration of vertical nanosheets is observed in low-energy Ar$^+$ and Xe$^+$ irradiated samples. An appreciably high magnetization value of 1.7 emu/g was observed for edge-oriented nanostructured pristine MoS$_2$ thin films, which decreased after ion irradiation. From X-ray photoelectron spectroscopy (XPS) data, it is evident that, due to oxygen incorporation in the sulfur vacancy sites, Mo5+ and 6+ states increase after ion irradiation. The density functional theory (DFT) calculations suggest that the edge-oriented spins of the prismatic edges of the vertical nanosheets are primarily responsible for the high magnetic moment in the pristine film, and the edge degradation and reduction in sulfur vacancies by the incorporation of oxygen upon irradiation result in a decrease in the magnetic moment.

* Corresponding author: santanu1@physics.iitd.ac.in

## 1. Introduction:

Molybdenum disulfides ($MoS_2$), among the 2D transition metal dichalcogenides (TMDC), are intriguing materials with different interesting characteristics such as effective catalytic [1], photoconductive [2], physicochemical [3], and lubricant properties [4], etc. It has a layered structure, a thickness-dependent bandgap (1.9 eV direct band gap for monolayer and 1.2 eV indirect bandgap for bulk), inter-layer weak Van der Waals interaction, and an intra-layer strong covalent bond [5], and strong spin-orbit coupling [6]. Ferromagnetic semiconductors are the key elements for the development of spintronics devices [7]. Semiconductors are often diamagnetic at room temperature. To induce ferromagnetism, researchers have attempted to dope semiconductors with ferromagnetic materials. Although promising results have been seen in this process, including room temperature ferromagnetism (RTFM), solubility is always an issue, which hinders the application of these materials in spintronics [8–11].

Defects are omnipresent in solids. A significant density of defects in solids can alter their electrical, optical, and magnetic properties [12]. Spintronics applications will greatly benefit if any non-magnetic semiconductor behaves like a magnetic semiconductor as a result of defects since there won't be a solubility issue [13–16]. Depending on the growth process, $MoS_2$ thin film exhibits edge-terminated nanostructure and sulfur vacancies. Bulk $MoS_2$ is a diamagnetic semiconductor, but as edge-terminated atoms have different coordination than bulk atoms, it shows a non-uniform spin distribution, different reconstruction, and unique magnetic properties compared to bulk films [17]. In this work, we have shown a comparatively high magnetic moment at room temperature of nanosheets of $MoS_2$ primarily contributed from the edge states.

To create defects and change structural and other physical properties, low-energy ion irradiation is a unique tool, as one can control ion type, ion irradiation energy, doses, etc. In low-energy regimes, ions transfer their energy through elastic collisions, known as nuclear energy loss ($S_n$), which causes various point defects, including vacancies and interstitials, as a result of the collision cascade. Few ion irradiation-induced magnetism in the $MoS_2$ monolayer [14], a few layers [13,18], and bulk [15] have been reported. Isolated vacancies, vacancy clusters, generation and reconstruction of edge states, defects, lattice distortion, and reconstruction, etc. have been mentioned as the reasons for the induced RTFM. In this work, defects were created in the samples by keV ion irradiation to understand the impact of defects on the observed magnetic moment. A comparative study of the heavy and light mass ion irradiation effects on the magnetic properties has also been done.

In order to understand this, CVD-grown nanostructured $MoS_2$ thin films are irradiated by 40keV $Ar^+$ and 100keV $Xe^+$ at different fluences to observe the effect of irradiation on the magnetic properties of these films. The description above makes it rather evident that $S_n$ plays a key role in producing the defects, and it is dependent on both the incident ion's mass and energy. Treating $MoS_2$ thin films with low energy and comparatively heavy ions became highly intriguing to create defects on a wide scale. The tuning of ferromagnetism in the samples is explained as an interplay between edge state and irradiation-induced defects through different experimental studies and DFT-based calculations. The interplay between edge state effect and defect in the variation of RTFM in nanostructured $MoS_2$ is rarely available in the experiment-based literature.

## 2. Experimental and computational details:

### 2.1 Experimental:

Chemical vapor deposition (CVD) has been proven to be a predictable process to grow high-quality, 2D-layered TMDC material on a large scale [5,19,20]. Here, large-area, high-quality, and pure-phase $MoS_2$ nanostructured thin films are grown on a $SiO_2$-coated Si substrate by thermal CVD. Fig. 1(a) shows the schematic diagram for the thermal CVD system for the growth of the films. In most cases, a multi-zone or double-zone furnace is used for the $MoS_2$ growth; however, growth was accomplished in a single-zone furnace in the present work. All the parameters, such as precursor quantity, gas flow rate, distances between precursors and substrate, temperature, and substrate type, were optimized according to a single-zone furnace. Sulfur (S,150 mg) and molybdenum trioxide ($MoO_3$,15 mg) powder were used as precursors. $MoO_3$ powder was kept in the central zone of the furnace, where a temperature of ~850°C was maintained, and S powder was kept at ~120°C in the quartz tube of length 90cm and diameter 3cm inside the furnace, so that both were evaporated at the same time. The furnace was heated up to 850°C at a ramp rate of 12°C per minute, and the deposition time was 20 minutes. The deposited film was allowed to cool naturally. When the cleaned $SiO_2$/Si substrate was kept face up and 12 cm away (~ 550 °C) from the $MoO_3$ powder (shown in Fig. 1(a)), a pure-phase vertical nanostructured $MoS_2$ thin film was obtained on a large area. The substrate height is crucial here because it has to be maintained in the region of intensive gas flow. The substrate was kept at 1.2 cm from the bottom in this case. As the distance between S and $MoO_3$ powder is 26 cm, a slightly higher gas flow rate was needed. 50 sccm of Ar gas flow was maintained in this case using a mass flow controller. Ar gas flowed for 30 minutes prior to the deposition to establish an inert environment to prevent contaminants. At first, S vaporized in the upstream region and flowed with the Ar gas, reaching the central part where $MoO_3$ started to vaporize. Sulfur reacts with $MoO_3$ and forms $MoO_2$, as the quantity of S is higher than the $MoO_3$ powder, then it reacts with $MoO_2$ and forms $MoOS_2$, and finally excess S reacts with $MoOS_2$ to form $MoS_2$ [5,16,21,22]. The related chemical reactions are:

1. $2MoO_3 + S \rightarrow MoO_2 + SO_2 \uparrow$

2. $2MoO_2 + 5S \rightarrow 2MoOS_2 + SO_2 \uparrow$

3. $2MoOS_2 + S \rightarrow 2MoS_2 + SO_2 \uparrow$

Subsequently, CVD-grown nanostructured thin films are irradiated with inert 40 keV $Ar^+$ and comparatively heavier 100 keV $Xe^+$ with different irradiation doses. The nuclear and electronic stopping power ($S_n$ and $S_e$) and the ion ranges (R) are calculated for both the ions by the Stopping and Range of Ions in Matter (SRIM) simulation software [23–25]. For 40keV, the Ar ion ranges with straggling, $S_n$, and $S_e$ are 29±15.8nm, 0.76, and 0.28 keV/nm, and for 100keV, the Xe ion ranges with straggling, $S_n$, and $S_e$ are 30±11.3nm, 3.16, and 0.37 keV/nm respectively (See Fig. S1 and Fig. S2 of Supporting Information). Hereafter, the pristine, 40keV $Ar^+$ irradiated samples with irradiation fluences 1e13, 5e13, 1e14, 5e14, and 1e15 ions/cm$^2$ will be denoted by A0, A1, A2, A3, A4, and A5. The 100 keV $Xe^+$ irradiated samples with irradiation fluences 1e13, 5e13, and 1e14 ions/cm$^2$ will be denoted by X1, X2, and X3, respectively.

X-ray diffraction (XRD) measurements were carried out on all the samples to identify the crystalline phase. The XRD pattern was taken in typical grazing incidence XRD mode (≤1º) by PANalytical X'Pert$^3$ in the 2θ range from 10º-60º with CuK$_\alpha$ radiation ( λ=1.5406A°). Raman spectroscopy was done at room temperature by the Renishaw Micro Raman Spectroscope to determine all the active molecular vibrations. A diode laser with a 532nm wavelength was used as a source. Field emission scanning electron microscopy (FESEM) was done by secondary electron mode (TESCAN instruments) to know the surface morphology and also the surface modification, and an Electron probe micro-analyzer (EPMA) (SHIMADZU EPMA-1720) to check the purity of the samples. X-ray photoelectron spectroscopy (XPS) spectra were recorded using AXIS Supra with the monochromatic Al Kα X-ray source (1486.6 eV). The pass energy used was 20 eV, and the overall resolution was ~ 0.6 eV. The pressure during the measurements was ~3×10$^{-9}$ mbar. Magnetic measurements were done by MPMS3 magnetometer (Quantum Design) at room temperature and also at 2K.

**2.2 Computational details:**

The Vienna *ab initio* simulation package (VASP) [26] with projected augmented wave (PAW) [27] potential was implemented for first-principles-based calculations under the framework of DFT [28,29]. The exchange-correlation interactions among electrons were accounted for by the generalized gradient approximation (GGA) with the functional form as proposed by Perdew-Berke-Ernzerhof (PBE) [30]. The structures were relaxed until the Hellmann-Feynmann forces on each atom were less than $10^{-5}$ eV/Å. The Γ-centered 1×4×1 k-grid was used to sample the Brillouin zones. The electronic self-consistency loop convergence was set to $10^{-5}$ eV, and the kinetic energy cutoff was set to 520 eV for plane-wave basis set expansion. All the edges were built with a vacuum of 12 Å (8.7 Å) in the z(x) direction to avoid electrostatic interactions among the periodic images. The two-body Tkatchenko-Scheffler Van der Waals scheme was employed for obtaining optimized structures [31,32].

**3. Results and discussion:**

Fig. 2 shows the XRD patterns of the pristine, Ar$^+$, and Xe$^+$ irradiated MoS$_2$ samples grown on SiO$_2$/Si substrates at room temperature. From the XRD patterns of MoS$_2$ nanostructured thin films, the formation of a pure 2-H phase [13] with trigonal prismatic coordination geometry with space group P6$_3$/mmc [33] is confirmed. The peak is situated at ∼14.34° is corresponding to the (002) plane [13,34] of MoS$_2$. As it shows only one peak, it is concluded that it has oriented growth in the (002) plane and has (002) edge sites [35]. No other phases or impurities are observed after irradiation in the XRD pattern. From Fig. 2(a), it is evident that the intensity of the peak decreases and the FWHM increases with increasing irradiation fluences from 0.76° to 1.25° from A0 to A5 samples. As FWHM increases, it confirms the deterioration or amorphization of the lattice after ion beam irradiation. There is a left shift of the (002) peak from 14.34° to 13.91° after ion irradiation, which indicates that after ion irradiation, strain is induced [36,37]. From Fig. 2(b), it is also clearly seen that after Xe ion irradiation, there is also a decrease in intensity of the (002) peak, and FWHM increases with increasing ion irradiation fluences, indicating degradation in crystallinity, and shifting of the (002) peak to the lower Bragg angle indicates induced strain after Xe$^+$ irradiation [38]. The (002) peak position and FWHM for all the samples are tabulated in Table 1. After ion irradiation average crystallite size decreases for both cases as lattice degradation and amorphization occurred (See supporting information).

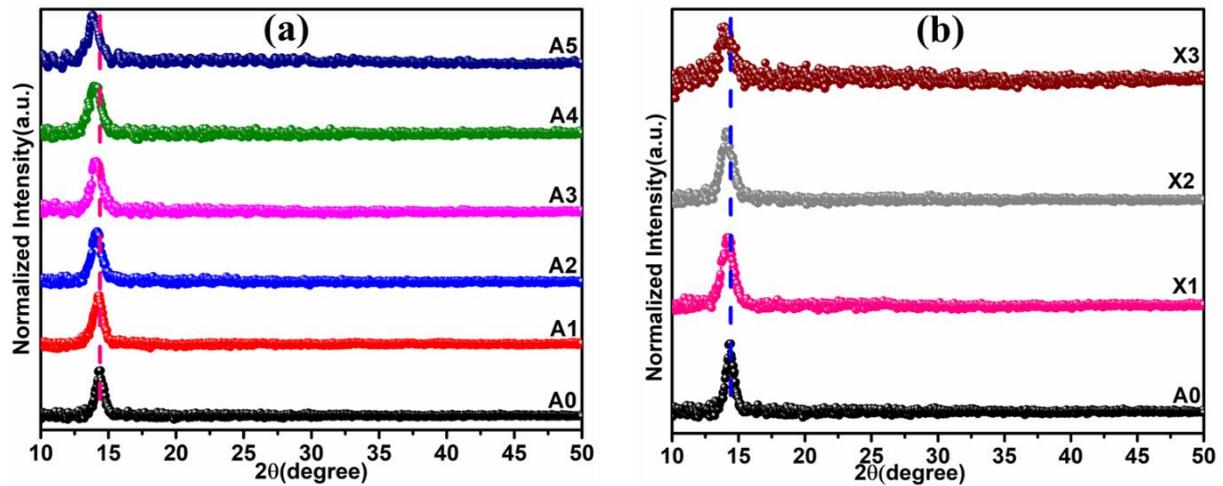

Fig. 2. XRD pattern of (a) pristine and Ar$^+$ irradiated samples. (b) pristine and Xe$^+$ irradiated MoS$_2$ samples. For both cases, there is a shift towards the lower Bragg angle in the position of the (002) peak and an increase in FWHM with increasing irradiation fluences.

Fig. 3 shows the Raman spectra of the pristine, Ar$^+$, and Xe$^+$ irradiated samples at room temperature. The peak situated at ∼383.5 cm$^{-1}$ is denoted by $E_{2g}^1$ peak (in-plane vibration of Mo and S atoms), and at ∼ 409 cm$^{-1}$ is denoted by A$_{1g}$ peak (out-of-plane vibration of S atoms). The peak situated at ∼452 cm$^{-1}$ is assigned to the second-order longitudinal acoustic (2LA) mode [18,39]. From Fig. 3(a) and (b), it is evident that for both cases (after Ar$^+$ and Xe$^+$ irradiation), with increasing ion irradiation fluences, the intensity of the peaks decreases and FWHM increases, which confirms degradation in crystallinity after irradiation. The increase in the FWHM indicates the formation of defects (like vacancies and vacancy clusters). The $E_{2g}^1$ peak is more sensitive (clear left shift shown in Fig. 3(a) and (b)) than peak A$_{1g}$ for irradiation as the covalent bond between two nearby Mo atoms was decreased due to the presence of S-vacancies, leading to a more pronounced decrease in the in-plane vibration energy compared to the out-of-plane vibration [40,41]. The reasons behind the magnetism in such nanostructured MoS$_2$ thin films are mainly sulfur vacancies and edge-terminated structures [15,18]. As the A$_{1g}$ peak denotes out-of-plane vibration of S atoms, the A$_{1g}$ /$E_{2g}^1$ peak ratio tells about the presence of edge-terminated structures [42]. This ratio decreases in both Ar$^+$ and Xe$^+$ irradiation cases with increasing irradiation fluences (see Table S1a of supporting information). As the saturation magnetization also decreases with increasing irradiation fluences, there is a one-to-one correspondence between edge-terminated structures and the magnetization.

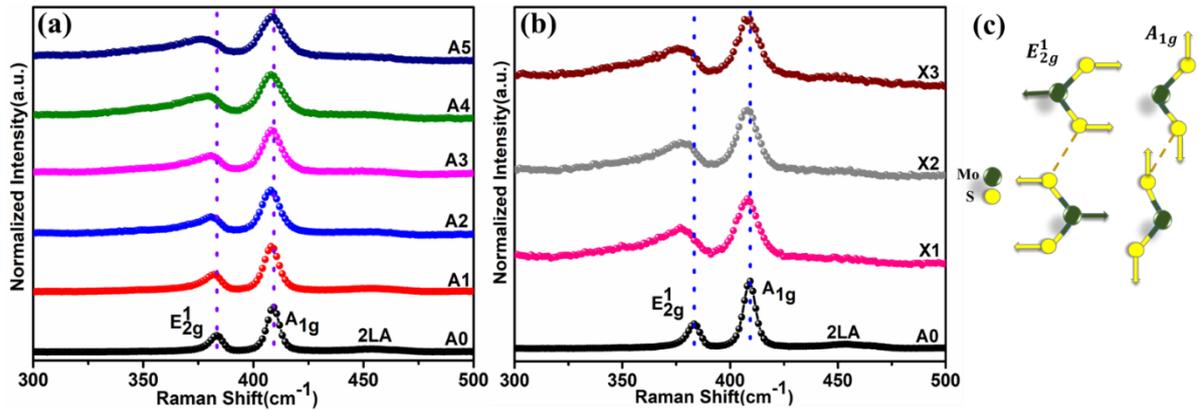

Fig.3. Raman spectra of (a) pristine and Ar$^+$ irradiated samples. (b) pristine and Xe$^+$ irradiated MoS$_2$ samples. (c) schematic diagram of in-plane and out-of-plane vibration of S and Mo atoms.

Fig. 4 shows the FESEM images of as-synthesized and Ar$^+$ and Xe$^+$ irradiated nanostructured MoS$_2$ samples. It is evident from Fig. 4(a) that the pristine sample contains many vertical nanosheets with edge termination. It has been seen that after Ar$^+$ irradiation, there is a deterioration in the vertical nanosheets. Fig. 4(c) and (d) show that there are more horizontal nanosheets and that the vertical nanosheets are thicker than the pristine sample, respectively (See Fig. S3 in supporting information). As unsaturated spin at the prismatic edges is the main reason behind the magnetism in nanostructured MoS$_2$ thin films, thick films contain a smaller number of prismatic edges due to the secondary nucleation of these edges [17]. It is obvious that in this case, induced magnetism is dependent on the surface edge states. Fig. 4(e) and (f) make it abundantly obvious that at higher irradiation fluences, horizontal nanosheets also degrade in addition to the damage done to vertical nanosheets. Similar events also occurred with Xe ion irradiation; however, since Xe ion is heavier than Ar ion and has high nuclear energy loss, comparatively lower fluences of Xe ion destroyed vertical nanosheets. In this case, the pristine film exhibits the highest magnetization, which subsequently decreases (discussed later), and additionally, FESEM images demonstrate that the edge-terminated nanostructure deteriorates with increasing ion fluences. So, it strongly suggests that magnetization is highly controlled by this edge-terminated nanostructure.

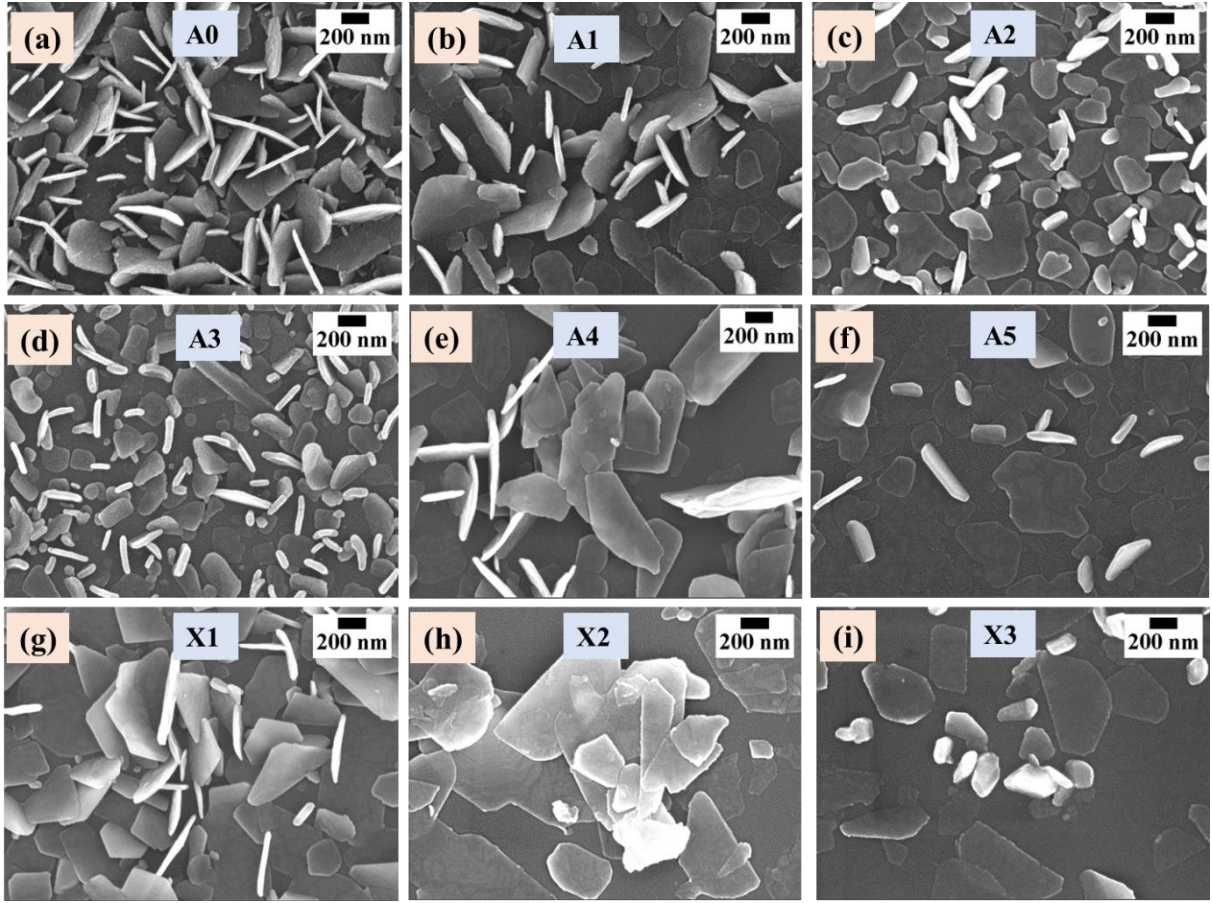

Fig. 4. FESEM images of (a)-(f) pristine and Ar$^+$ irradiated samples and (g)-(i) Xe$^+$ irradiated samples. It is clear from the figures that for both cases the vertical nanostructures are degraded.

EPMA analysis is carried out to show the distribution of Mo and S over the sample and also to confirm the absence of any magnetic impurities [43]. Here, mapping and qualitative analysis are performed for the A0, A2, and X2 samples. Mapping and qualitative analysis are shown for the A0 sample in Fig. 5 on two different points (see Fig. S4 and S5 in Supporting Information for A2 and X2 samples), and it confirms the presence of only Mo and S in all the samples. The Si signal is from the substrate and has no magnetic contribution. At first, an area is selected, which is shown in Fig. 5(a), and mapping is carried out independently for Mo and S elements. For Mo and S detection, a PET (Pentaerythritol) crystal detector is used. Fig. 5(b) shows the mapping for S, and Fig. 5(c) shows the mapping for Mo. Fig. 5(d), and (e) show the qualitative analysis for Mo and S at two different points. From Table S1.b, it is evident that after ion irradiation, the relative S/Mo ratio decreases with increasing ion fluences, confirming the creation of S vacancies.

The Magnetization versus magnetic field (M versus H) loops as obtained from the superconducting quantum interference device (SQUID) measurements at room temperature for all the samples are shown in Fig. 6. The measurements carried out at 2K are shown in Fig. S6.

The sample was placed parallel to the magnetic field, and the substrate contribution was subtracted from the total signal. The applied magnetic field varies from 0 to 2T at 300K and 2K. A0 to A4 samples show a hysteresis loop at room temperature and also at 2K, but at higher fluences, they behave as a diamagnetic sample. The zoom version of the central part is shown in the inset of Fig. 6. The pristine sample shows the highest magnetization value of 1.7 emu/g at room temperature and decreases with increasing irradiation fluences in both $Ar^+$ and $Xe^+$ irradiated samples. The values of saturation magnetization ($M_s$), coercive field ($H_c$), and remnant magnetization ($M_r$) at 300K are tabulated in Table 2 and for 2K in Table S2. Magnetization versus temperature measurement is done in a standard oven set up from 300 to 900K at a 1000 Oe field for the pristine sample, which is shown in Fig. S7.b. It is evident from the M versus T curve that the transition temperature is 830K, i.e., nanostructured $MoS_2$ thin film is ferromagnetic at room temperature [17].

Transition metal-doped $MoS_2$ shows room-temperature ferromagnetism [8,10,11,44,45], but there is always a debate about whether the magnetism originates from the agglomeration of the transition metal or the synergetic effect of $MoS_2$ and dopants. Therefore, defect-induced magnetism was explored. The possible reasons for the induced magnetism in $MoS_2$ are isolated vacancies or vacancy clusters, edge states, and reconstruction of the lattices [15]. There are also some other reports where zigzag edges and defect-induced room-temperature ferromagnetism are induced in nanocrystalline $MoS_2$ films and nanoribbons [14,46]. There are few theoretical [47,48] and experimental studies [37,49] where strain tuning is correlated with defects-induced magnetism. Ferromagnetism was explained based on bound magnetic polaron (BMP) induced by the spin-interaction of trapped electrons of S-vacancies and Mo 4d ions [13]. Magnetic moments reported in different studies pertaining to $MoS_2$ are tabulated in Table S3. As can be seen, the reported magnetic moment values are all very low compared to the present case.

In the present work, nanostructured pristine $MoS_2$ thin films show RTFM with a moment of 1.7 emu/g. The unsaturated Mo atoms at the edge states and S vacancies are the main reasons behind the induced magnetism. The edge-terminated atoms have different stoichiometry than the bulk atoms, and Mo atoms will be unsaturated at the edges with octahedral coordination, which will result in spin polarization and induced ferromagnetism. The higher the number of unsaturated prismatic edges, the greater the enhancement in ferromagnetism. There is the formation of S vacancies, as evident from EPMA and XPS data (discussed later section) after ion irradiation. But as $MoS_2$ is a layered structured material, it tends to absorb oxygen at the surface, so after oxygen atom incorporation in the vacancy site, the magnetic moment decreases with increasing irradiation fluences [50,51], which is further clarified by XPS and theoretical analysis later. To confirm this experimentally, one pristine sample was annealed at 100°C for 2 hours in the presence of oxygen, and magnetic measurements were performed. It has been noticed that the magnetization value decreases from 1.7 emu/g to 0.29 emu/g for the pristine to $O_2$-annealed sample (See Fig. S7.a in supporting information). Hence, the decrease in magnetization on irradiation is qualitatively understood based on the following two processes (i) the incorporation of O into the S vacancies (as after ion irradiation vacancies are created, and O incorporated easily in these vacancies) and (ii) the deterioration of vertical nanosheets due to irradiation. The DFT calculations and XPS measurements presented in subsequent sections give further insight into the origin of magnetism.

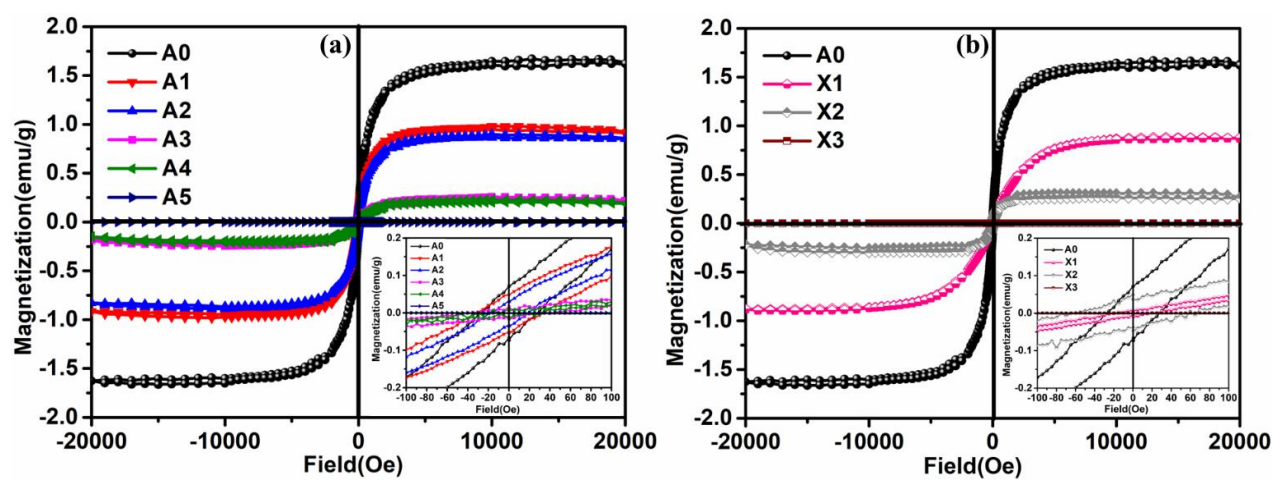

Fig. 6. (a) M versus H plot for pristine and $Ar^+$ irradiated samples at room temperature; (b) M versus H plot for pristine and $Xe^+$ irradiated samples at room temperature. A0 to A4 and A0 to X2 samples show hysteresis loop at room temperature, whether A5 and X3 samples show diamagnetic behaviour.

To clarify the reason behind the decrease in magnetization with increasing irradiation doses, the electronic structure of all the samples has been elaborately studied. No transition metal impurity is observed in the samples (see Fig. S8 in supporting information). Mo 3d and S 2p core-level XPS spectra of $MoS_2$ pristine and $Ar^+$ irradiated samples are shown in Fig. 7 in a stack. It is clear that after irradiation, there is a shift towards lower binding energy (BE) for all the peaks, and also peak broadening in the Mo 3d and S 2p core-level spectra, which correspond to disorder, defects, surface amorphization, and vacancy formation [52–54]. The shift towards lower BE indicates after ion irradiation there is formation of vacancies and defects.

In Fig. 8, the Mo 3d core-level XPS spectrum is deconvoluted into nine peaks with a pseudo-Voigt function by keeping the same FWHM for all the corresponding peaks. The background signals were subtracted by using the iterative Tougaard method by CasaXPS software and calibrated using the C 1s peak at 284.6 eV. To fit the Mo 3d spectrum, the separation and the relative area ratio between the two spin-orbit doublet 5/2 and 3/2 is maintained as 3.2eV and 1.5, respectively. As evident, $Mo^{4+}$ valence states predominate in all the samples, which confirms pure $MoS_2$ phase formation. $Mo^{4+}$ $3d_{5/2}$ and $3d_{3/2}$ are located at 229.6 and 232.8 eV, $Mo^{5+}$ $3d_{5/2}$ and $3d_{3/2}$ are located at 230.4 and 233.6 eV, and $Mo^{6+}$ $3d_{5/2}$ and $3d_{3/2}$ are located at 232.9 and 236.1 eV, respectively. Lower valence states of Mo are also present there, which are labeled as 1 and 2. The lower valence states are situated at 228.6 and 231.8 eV respectively [55–57]. The presence of lower valence states indicates there is some reduced Mo present i.e. there are creation of S vacancies and the formation of amorphized $MoS_{2-x}$ phases at the surface. So, the CVD-grown pristine and as well as the irradiated samples contain some defects and S-vacancies. The peak situated at 226.8 eV corresponds to the S 2s peak [58,59]. The $Mo^{4+}$ 3d5/2 state shifted from 229.6 eV to 229 eV and the lower valence state shifted from 228.6 to 228 eV from pristine to the highest irradiated samples, similar observation is shown by A. Santoni et. al [52]. With increasing irradiation doses, the intensity of peaks 1 and 2, and the peaks corresponding to Mo 5+ and 6+ states, is increasing, and the intensity of the peak attributed to S 2s is decreasing. An increase in the intensity of the peaks corresponding to Mo 5+ and 6+ states shows the incorporation of oxygen and its coordination with Mo, whereas at the same time, an increase in the intensity of peaks labeled as 1 and 2 and a decrease in the intensity of the peak ascribed to S 2s confirm that S vacancies are created after irradiation. The O1s core level XPS spectra show that the peak around 530.5 eV corresponding to lattice oxygen increases with increasing irradiation fluences (see Fig. S10 in supporting information), which also confirms the incorporation of oxygen in some of the sites created by S vacancies. The percentage concentration of all the states is tabulated in Table 3. The same trend is observed for $Xe^+$-irradiated samples. The Mo 3d and S 2p spectra (the peaks at 162.4 and 163.6 eV correspond to $S^{2-}$ 2p 3/2 and 1/2, respectively, and the spin-orbit separation is 1.2eV) of the X1, X2, and X3 samples are shown in Fig. S9.

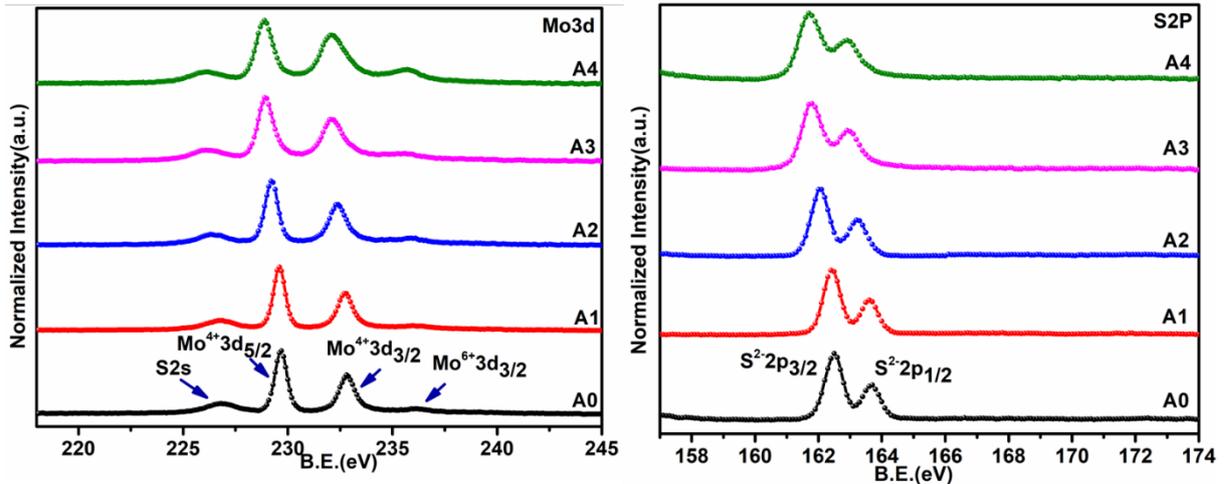

Fig. 7. Mo 3d and S 2p core level XPS spectra of pristine and Ar$^+$ irradiated samples. There is a shift towards lower binding energy in both Mo 3d and S 2p spectra.

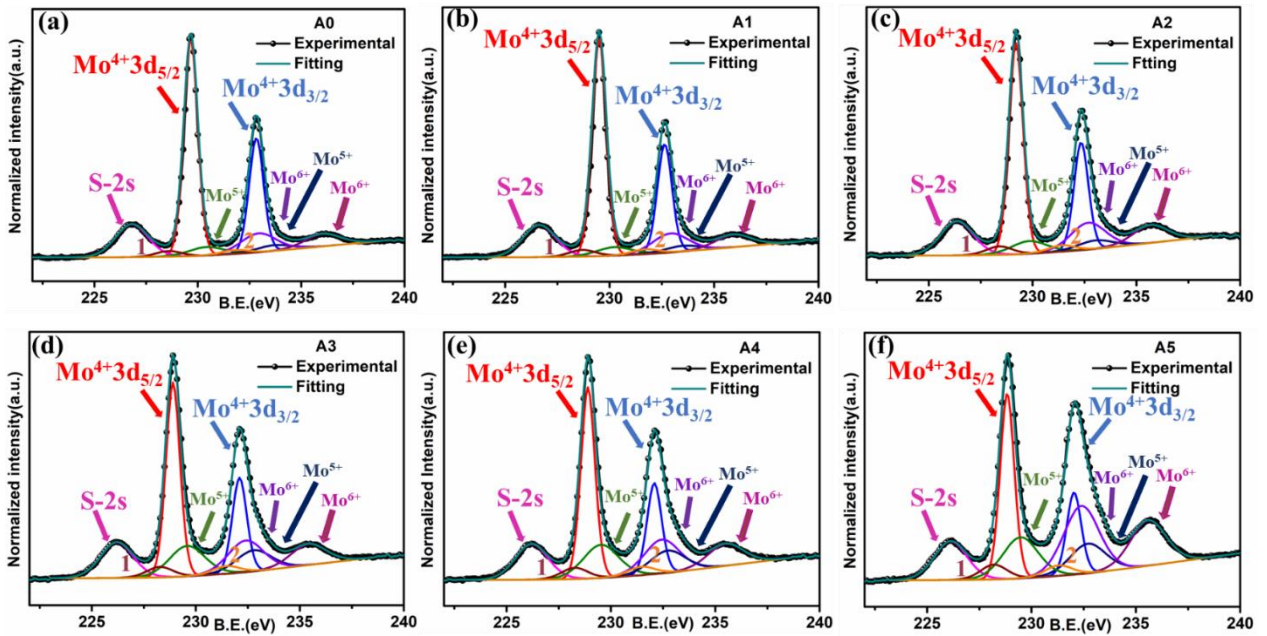

Fig. 8. Mo 3d core level XPS spectra of (a) pristine and (b)-(f) Ar$^+$ irradiated samples. The intensities of peaks corresponding to lower and higher valence states increase with increasing irradiation fluences.

To support the rationale behind the induced magnetism and to corroborate the experimental findings, first-principles-based theoretical calculations on layered MoS$_2$ were conducted. The structural information of bulk MoS$_2$ along with S vacancy and O adsorption at the S vacancy site are shown in Fig. S12. The magnetic moment in all three cases was calculated to be 0.0 $\mu_B$, confirming its nonmagnetic behaviour in the bulk. Further, the analysis focused on edge-oriented MoS$_2$, where a magnetic moment of 3.2 $\mu_B$ was observed. This magnetism arises due to the different coordination of edge-terminated atoms than that of the bulk atoms. Furthermore, as S vacancies serve as the basic reason to induce magnetism in $MoS_2$, calculations were directed toward accessing the effect of S vacancies. The magnetic moment of $MoS_2$ with an S vacancy at the edge was found to be 3.85 $\mu_B$. This increase in magnetic moment upon the introduction of the S vacancy aligns with findings reported by A.K. Anbalagan et al., providing additional validation for this phenomenon. Moreover, upon ion irradiation, the magnetic moment is decreasing due to O incorporation at S vacancy positions for nanostructured $MoS_2$. The O-incorporated S vacancy showed a drop in magnetic moment in nanostructured $MoS_2$, the same as in the case of the experiment, and was observed to be 3.62 $\mu_B$. In a word, it is concluded that our experimental findings and DFT-based calculations are in accordance with each other.

## 4. Conclusions:

$MoS_2$ nanostructured thin films were prepared using the CVD technique and subsequently irradiated by Ar and Xe ions. The nanostructured, pristine sample shows room-temperature ferromagnetism with a saturation magnetization value of 1.7 emu/g. Low-energy $Ar^+$ and $Xe^+$ irradiation is performed on these CVD-grown samples to create vacancies and defects and to tune the magnetic properties. XRD and Raman spectroscopy analysis confirmed the deterioration in crystallinity after irradiation. FESEM images confirmed edge-terminated nanostructure formation and deterioration of these structures after ion irradiation. EPMA analysis clearly showed there was no magnetic impurity present in the samples. The decrease in magnetic moment on irradiation has been explained on the basis of the deterioration of edge states and the incorporation of oxygen in S vacancies from XPS analysis. The present study suggests that the presence of edge states is more important for observing ferromagnetism in nanostructured $MoS_2$, as deterioration in edge-terminated nanostructure and incorporation of oxygen at S vacancy sites lead to a decrease in magnetic moment even when there is an increase in S vacancies.

**Acknowledgments:**

The authors are grateful to the Physics Department at IIT Delhi for the XRD, Raman, and SQUID facilities and the Central Research Facilities (CRF) for the FESEM, EPMA, and XPS facilities. The authors also acknowledge the low-energy ion beam facilities provided by IUAC, New Delhi. S. Dey and A. Phutela, acknowledge IIT Delhi for the senior research fellowship and S. Bhattacharya is thankful to the Science and Engineering Research Board (SERB) for the financial support under the core research grant (grant no. CRG/2019/000647).